\documentclass[aps,prb,twocolumn,twoside,a4paper,showpacs]{revtex4}
\usepackage{graphicx} 
\begin{document}

\title{Liquid-liquid coexistence in the
phase diagram of a fluid confined \\ in fractal porous materials.}
\author{V. De Grandis}  
\author{P.~Gallo}\author{M.~Rovere}\email[Author to whom correspondence 
should be addressed: ]{rovere@fis.uniroma3.it}

\affiliation{                    
  Dipartimento di Fisica, 
Universit\`a ``Roma Tre'', \\ and Democritos National Simulation Center, 
INFM-CNR,\\
Via della Vasca Navale 84, 00146 Roma, Italy.
}

\begin{abstract}
Multicanonical ensemble sampling simulations have been performed
to calculate the phase diagram of a Lennard-Jones fluid
embedded in a fractal random matrix generated through 
diffusion limited cluster aggregation. The study of the
system at increasing size and constant porosity shows that
the results are independent from the matrix
realization but not from the size effects.
A gas-liquid transition shifted with respect to bulk is found.
On growing the size of the system
on the high density side of the gas-liquid 
coexistence curve it appears
a second coexistence region between
two liquid phases. These two phases are characterized by a different
behaviour of the local density inside the interconnected
porous structure at the same temperature and chemical potential.

\end{abstract}


\maketitle

\section{Introduction}
The effect of confinement on the phase diagram
of fluids is a longstanding problem with a
wide interest from the theoretical point of view 
and a wide area of possible applications
such as catalysis, adsorption and filtration. 
A large phenomenology on this subject can be
found~\cite{gelb}.
Real porous materials show generally a complex structure made of 
an interconnected network of pores. Vycor glass and silica gels
are relevant examples. 
It is important to distinguish
between two classes of systems.
In hosting media 
with very high porosity like silica aerogel ($90\%-98\%$) it
has experimentally been observed that the gas-liquid transition is preserved
although the confinement causes
a reduction of the critical temperature and 
a shrinkage of the gas-liquid coexistence curve~\cite{wong1,wong2}.
In mesoporous materials with 
a porosity in the range between $30\%$ and $60\%$, as for instance
Vycor glass, there are not 
direct observations of equilibrium 
phase transitions~\cite{monette}. This last behaviour is predicted by
theoretical work performed in the framework of lattice 
gas models~\cite{gelb,monette,kierlik,sarkisov1,woo-monson,detche1,detche2}.
This type of approach is connected to the more general issue
of the effects of quenched disorder on 
critical phenomena~\cite{degennes,imry,fisher,maritan,toigo}.
In dilute silica aerogels with high porosity, however, 
phase transitions between equilibrium phases cannot be 
theoretically excluded~\cite{rosinberg,krakoviack}
although more recent experiments~\cite{beamish,wolf}
and mean field calculations~\cite{detche1,detche2} 
questioned the early experimental findings
of a gas-liquid transition in $^4He$ in aerogel.

For dilute silica aerogels off lattice liquid state models have been 
introduced~\cite{madden_glandt,page_monson}
to take into account more in detail the microstructure
of the confining systems. 
In these off lattice models the disordered porous material is 
modeled as a system of spheres frozen in a predefined
structure. The quenched matrix approximately accounts for the
geometric constraints of the interconnected
pore structure of the solid. 
From both computer simulations~\cite{page_monson,sarkisov} and
integral equation methods~\cite{rosinberg} it has been found
that the phase diagram of these models can show, besides 
the gas-liquid coexistence (GLC),
the presence of a second transition.
This transition can be 
a gas-gas or a liquid-liquid coexistence depending on the
liquid-matrix type of interaction. 
Nonetheless the appearance of this coexistence 
and its details depend  
on the way in which the confining 
random matrix is realized~\cite{alvarez}.
Moreover relevant wetting or drying effects could
be present~\cite{page_monson}. 

Therefore while
the coexistence of different fluid forms
in a simple monatomic fluid confined
in silica aerogels is of uttermost interest, its 
existence is still a
controversial result. 
The fact that the phase diagram has shown to depend so much on the
details of the quenched random matrix 
has precluded so far more detailed microscopic analysis 
on these confined fluids and a systematic study of size effects.

From these considerations and a more 
recent theoretical analysis based on 
integral equation methods~\cite{krakoviack} 
it emerges the importance of realizing by computer simulation 
an host system which accounts microscopically for
the high porosity and the fractal structure of the silica aerogel.
Its structure is due to the process
of formation of the gel through the random aggregation
of the silica particles and it is
measured in small angle
scattering experiments
where a fractal dimension of $D=1.8$ has 
been estimated~\cite{hasmy1,hasmy2,hasmy3}.
These microscopic features are not present in any of the previous
lattice or off lattice simulated systems, mentioned above.

In order to build up a confining medium with the 
structure of silica aerogels,
we have implemented a numerical procedure~\cite{degrandis}
based on the diffusion limited cluster-cluster aggregation
(DLCA)~\cite{hasmy1,hasmy2,hasmy3}.
It has been shown~\cite{hasmy1,hasmy2,hasmy3} 
that the DLCA algorithm is able to reproduce the 
structure factor and the fractal
dimension of silica aerogels. In a previous study 
we calculated with the Multicanonical Ensemble 
Sampling (MES)~\cite{berg,wilding1,wilding2} 
the phase diagram of a Lennard-Jones fluid confined in the matrix
generated with the DLCA
and we found a GLC curve~\cite{degrandis} and no evidence of a 
second transition except for a possible signature in 
a shouldering on the liquid
side of the coexistence curve of a liquid-liquid coexistence (LLC).
As it is well known for aerogels~\cite{hasmy2} 
systems with the same porosity show the same average size of the clusters,
the same connectivity, the same behaviour and range of the fractal scale.
We expect therefore that performing simulations 
with confining media at constant porosity and increasing size
that the behaviour of the liquid will be independent from the details
of the realization of the matrix apart for the size effects 
themselves since the liquid is embedded in self similar structures.

In this paper we report the results obtained by increasing the
size of the confining medium
at constant porosity with the aim of inquiring about the
existence of a LLC in the Lennard-Jones simple fluid possibly 
hidden in our past study
by the small size of the simulated system.

\section{Simulation details}
To generate the off lattice matrix 
in a simulation box of side $L$
we introduce a number $N_s$ of spheres 
of diameter $\sigma_s$ randomly placed and apply
periodic boundary conditions.    
After the application of the DLCA algorithm 
the final configuration of the matrix consists of a percolated cluster
of $N_s$ spheres. For our former system, which we will
call in the following system I~\cite{degrandis},
we used $L=15 \sigma_s$ and 
$N_s=515$ in order to have a volume fraction $\eta = 0.08$ corresponding to
a porosity $P=92 \%$. For the system simulated in the present work,
system II, we fix
the same $\eta$ and porosity increasing the box length to $L=20 \sigma_s$.
In this case the system consists of $N_s=1222$ spheres.
The average cluster size is $4 \sigma_s$ independent from the system size. 
The particles of the confined fluid interact with a Lennard-Jones potential
with $\sigma=\sigma_s$ at variance with real systems where
the aerogel spheres are much larger than the fluid particles. 
In the following Lennard-Jones units
will be used. The potential is truncated at $r_c=2.5$.
The interaction
between the matrix spheres and the fluid particles is represented
by a hard sphere potential with diameter $\sigma$.
The simulation of the phase diagram of the confined fluid
is performed by Monte Carlo in the grand canonical ensemble (GCMC)
with the algorithm introduced by Wilding~\cite{wilding1,wilding2}.
By varying the chemical potential $\mu$ at constant temperature $T$
we follow the behaviour of the density fluctuations and calculate
the density distribution functions (DDF) $P(\rho)$. 
In a two phase region the DDF develop a double peak structure.
The coexistence point between the two phases is located with
the Wilding criterion~\cite{wilding2} by tuning the chemical potential
at constant temperature to obtain equal area under the two peaks.
In the subcritical region where a large free
energy barrier is present between the two phases 
the MES~\cite{berg,wilding1,wilding2}
has been implemented to enhance the sampling of the two 
coexisting phases.  
 
From the double peaked DDF we can extrapolate the
biased sampling function at different chemical potentials and
temperatures by means of the histogram reweighting technique~\cite{ferrenberg}.
With the bias sampling function the MES can be carried on. In the
final step the bias must be subtracted to recover the real
distribution function~\cite{degrandis}.
To equilibrate system II $10^8$ Monte Carlo moves have
been generated, while equilibrium properties have been calculated
with $10^9 - 10^{10}$ Monte Carlo moves. The system has been 
investigated in the range of temperatures from $T=0.95$ to $T=0.825$.  

\section{Density distribution functions and phase diagram}
\begin{figure}[ht]
\begin{center}
\includegraphics[width=80 mm]{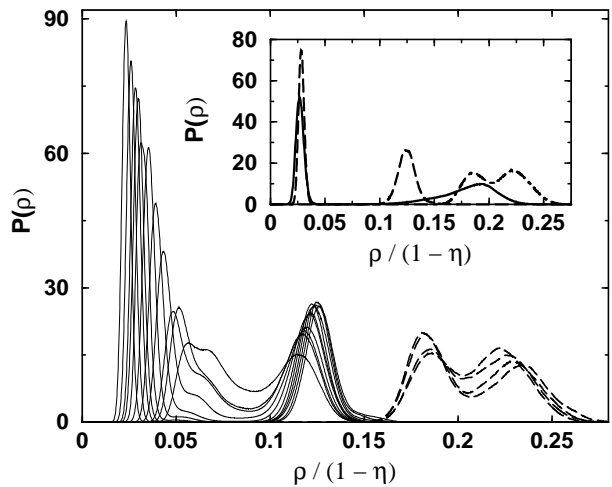}  
\caption{Density distribution functions (DDF) of the fluid
at the gas-liquid coexistence (solid line)
for system II in the range of
temperatures below the critical point from $T=0.825$ to $T=0.95$
and liquid-liquid coexistence (long-dashed lines) 
in the range of temperatures from $T=0.825$ to $T=0.85$.
DDF with closer peaks correspond to higher temperatures. 
In the inset 
$P(\rho)$ of the liquid-gas coexistence in system I
(solid line) compared 
with the liquid-gas (long-dashed line) and the liquid-liquid
(dot-dashed line) distribution functions of system II at $T=0.85$.}
\protect\label{fig:1}
\end{center}
\end{figure}

In Fig.~\ref{fig:1} the bimodal DDF of system II are reported
in the thermodynamical range where the MES analysis has been performed.
The low density peaks corresponding to the gas phase show a behaviour
very similar to the corresponding quantities in system I
therefore the gas
branch of the coexistence curve appears as almost independent from the
size of the system. On the liquid side coexisting with the gas
we observe instead large
differences in the structure of the peaks. In system I they were 
found to be broad and asymmetric~\cite{degrandis}. 
Now in system II the form of
the peaks is symmetric and sharper.
To be more precise for temperatures $T<0.90$ the 
density fluctuations are of the order of $0.07$ in system I and
$0.006$ in system II in the same range of temperature 
variation $\Delta T \simeq 0.075$.
By increasing the chemical potential at values above the
liquid-gas coexistence we find in the adsorption isotherms
the appearance of a second coexistence in system II. 
The corresponding DDF are also reported in 
Fig.~\ref{fig:1} where they show very well defined peaks. 
In the inset of Fig.~\ref{fig:1} it is shown how 
the bimodal $P(\rho)$ in the system I at $T=0.85$ is modified 
in system II with the emerging of new structures.
While the low density peak corresponding to the gas phase
remains in the same position in system II, the liquid peak
of system I splits in two in the larger simulation box
disclosing the presence of a possible LLC.
The new bimodal structure is obtained by increasing
of $2 \%$ the value of the chemical potential from 
the value at the gas-liquid coexistence. 
The size of system I was not enough large to allow
the resolution of the different contributions to the
density fluctuations in the system in the region of the
liquid peak.

\begin{figure}[ht]
\begin{center}
\includegraphics[width=80 mm]{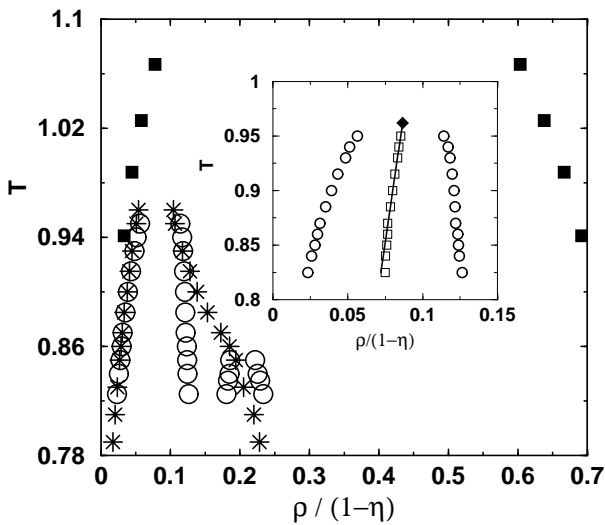}  
\caption{Phase diagram of system II (open circles)
compared with the one of system I (stars) and a portion
of the bulk liquid-gas coexistence curve (filled squares).
In the inset liquid-gas coexistence curve of system II (open circles)
with the diameter (open squares) and the best fit to the
rectilinear diameter law (solid line), 
the filled diamond represents the critical point
obtained by extrapolation of the fit.
}
\protect\label{fig:2}
\end{center}
\end{figure}

From the peak positions of the DDF with the Wilding criterion the 
coexistence curves of system II have been determined and reported
in Fig.~\ref{fig:2} where they are compared with the coexistence
curves of system I and with the bulk. 

The confinement causes a substantial
shrinkage of the gas-liquid coexistence curve and a shift toward
lower temperatures and densities of
the critical point.
The confinement also does not induce large changes with respect to
the gaseous phase of bulk fluid, in fact the gas branch
of the gas-liquid coexistence curve in confinement
almost coincides with the analogous branch in the bulk.
On the liquid side of system II a new
coexistence curve is now observed between
a low density liquid, 
which we shall call liquid I, and an
high density liquid, liquid II. 
We note that the range of extension of the
phase diagram of the confined system is not quite affected by
the system size. 
We checked that our results are independent from the matrix
realization by performing simulations with other two different
matrices generated with DLCA at the same porosity and $L=20$.
 
In system II, where
the GLC appears as a well defined curve,
it is possible to extrapolate the critical point position with
the use of the law of the rectilinear diameter and the
scaling law of the density~\cite{rowlinson}.
In the inset of Fig.~\ref{fig:2} we show the coexistence diameter
$\rho_d=(\rho_g+\rho_l)/2$.
The best fit to the rectilinear diameter law
$\rho_d = \rho_c + A(T_c-T)$
combined with the scaling law
$\rho_l-\rho_g = B(T_c-T)^\beta$
gives the following 
estimates of the liquid-gas critical parameters: $T_c=0.96$
and $\rho_c=0.0791$ to be compared with the
values for the bulk
$T_c=1.1876$ and $\rho_c=0.3197$~\cite{wilding1}. 
At variance with experimental results~\cite{wong1} we find a shift of the
critical point to a lower density, this is due to the assumption
of a pure repulsion between the fluid and the matrix~\cite{rosinberg}.
The exponent $\beta$ 
compatible with these critical values is 
in the range $0.30 \sim 0.32$ to be compared with the universal
exponent $\beta=0.3258$ found in the bulk~\cite{wilding1}. 
This also confirms the experimental finding that
the universal behaviour of the gas-liquid transition in the bulk
persists also in confinement~\cite{wong1,wong2}.     

\section{Liquid-liquid coexistence region}

\begin{figure}[ht]
\begin{center}
\includegraphics[width=80 mm]{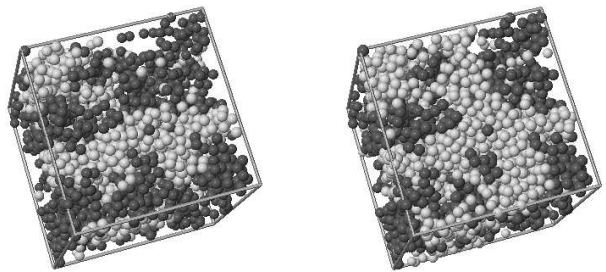}  
\caption{Snapshots of the configurations of the 
confined fluid at $T=0.85$ in the low density liquid
phase (liquid I) on the left and in the high 
density liquid phase (liquid II) on the right.
The gray and the black spheres represent the gel
and the liquid particles respectively. 
}
\protect\label{fig:3}
\end{center}
\end{figure}

In Fig.~\ref{fig:3} we report the snapshots of two coexisting
configurations of the fluid at $T=0.85$ in the region of
the second coexistence curve. 
From the snapshots we can observe that both liquid I and II are
characterised by the presence of a large cluster and a number of
smaller clusters with few particles.
We calculated 
the gyration radius $R_g$ of the larger cluster in the two liquid phases
in the range of temperature from $T=0.850$ to $T=0.825$.
$R_g$ is defined by $R_g=\sqrt{ \sum_i r^2_{ci}/N}$,
where $r_{ci}$ is the distance between the particle $i$ and
the center of mass of the cluster.
In liquid I we have an increase of number of particles in the system
upon increasing temperature and a
$3 \%$ increase in the number of particles 
corresponds to a $2.2 \%$ increase of the particles
in the cluster and to a $7.3 \%$ increase of the gyration radius.
In liquid II we observe the opposite trend
instead, an increase of number of particles is observed
upon decreasing temperature. Besides a
$5 \%$ increase in the number of particles 
corresponds to a $7 \%$ increase of the particles
in the cluster while $R_g$ grows only
for an $1.7 \%$.

\begin{figure}[ht]
\begin{center}
\includegraphics[width=75 mm]{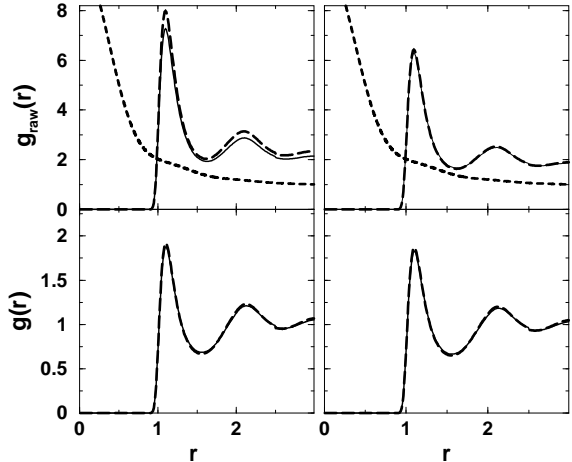}
\caption{Radial distribution functions of the fluid in the
two phases, liquid I on the left, liquid II on the right,
for T=0.850 (solid line), T=0.825 (long-dashed line).
In the upper panels the RDF 
as obtained from simulation are shown together with 
the excluded volume correction (dashed line).
In the lower panels the RDF are shown
after the correction is applied.}
\protect\label{fig:4}
\end{center}
\end{figure}

The radial distribution 
functions (RDF) of the fluid in the two
liquid phases (Fig.~\ref{fig:4}) have been also calculated 
with a canonical MC starting from
equilibrated configurations obtained from the GCMC.

The usual algorithm for obtaining the RDF from simulation
consists in normalizing the number of atom pairs in 
a spherical shell
with the uniform distribution of ideal particles 
in the same shell. Two of the $g_{raw}(r)$ functions obtained
with this procedure are reported as an example in the upper panels of
Fig.~\ref{fig:4} for the less dense liquid,
liquid I, and for the denser liquid, liquid II.
According to previous literature~\cite{soper}
when a fluid is confined 
the corrected RDF can be generated by
appropriately normalizing $g_{raw}(r)$
with the {\it uniform} radial distribution function $g_u(r)$
which depends on the geometry of the confining medium
and a factor $f=V/V_{eff}$ that accounts for the 
effective volume occupied by the particles:
$g(r)=g_{raw}(r) / ( f \cdot g_u(r) )$.
The function $g_u(r)$ can be determined with a convolution
of the structure factor of the aerogel system and the 
internal structure factor of the single hard sphere~\cite{soper}.
The resulting $g_u(r)$ is reported in the upper panels of Fig.~\ref{fig:4}.
The factor $f$ and $V_{eff}$ can be
empirically determined by imposing that
$g(r)$ goes to $1$ for large $r$. 
For liquid I
the factor $f$ decreases and $V_{eff}$ increases with 
increasing temperature, 
in the two cases reported in Fig.~\ref{fig:4},
$f=2.03$ for $T=0.850, N=1367$ and $f=2.18$ 
for $T=0.825, N=1329$.  
In the case of liquid II, in Fig.~\ref{fig:4}
are reported $T=0.850, N=1634$ and  $T=0.825, N=1717$, for which
we get approximately the same
value $f=1.81$, independent from the temperature.

\section{Conclusions}

With the use of the DLCA algorithm to build
a confinig structure with fractal character
we have found that the behaviour of the
confined fluid 
confined in a fractal structure
does not depend on the matrix realizations and 
finite size effects can be studied more systematically.
By increasing the size of the simulation box 
the gas-liquid transition is well characterized,
while a second coexistence curve appears on the
high density side of the gas-liquid binodal curve.
It is shown that
in the low density liquid phase (liquid I) the size of the liquid domains
increase with increasing temperature
and equivalently the effective volume occupied by the particles $V_{eff}$
grows. 
In the high density liquid phase (liquid II) instead there is no increase
of the size of the liquid domains and of the effective volume.
In this regime we have an increase of the local density which
is made possible by little rearrangements of the particles.
Even if drying effects due to the repulsive interaction between the fluid
and the substrate cannot be excluded it seems that 
the high porous open structure of the aerogel network,
built in our simulation, 
allows the formation
of an almost homogeneous liquid phase inside the free volume.

The interpretation of fluid adsorption phenomena in aerogels is 
still controversial.
In our simulation we studied equilibrium phase transitions and we
did not consider hysteretic behaviour. In this respect our treatment
is analogous to the integral equation study of
Krakoviack et al.~\cite{krakoviack}, where a gas-liquid transition
(but not a liquid-liquid one) 
is found for a fluid adsorbed in a high porosity aerogel. 
Calculations performed with mean field
density functional theory aimed to study nonequilibrium 
behavior show instead a different scenario. The disordered
character of the aerogel structure modifies the adsorption-desorption
mechanism although at high temperatures there are
indications of a gas-liquid transition~\cite{detche1,detche2}.
We need future experimental and theoretical 
work to get a more complete understanding of this
vaste and important phenomenology of fluid adsoprtion
phenomena. In this respect
computer simulation with complete finite size scaling analysis
seems to be necessary since we have shown here that
finite size effects are quite relevant.

\end{document}